\documentstyle[12pt,aasms4]{article}

\begin{document}

\lefthead{Levin}
\righthead{Runaway heating by r-modes}
\title{
Runaway Heating By R-Modes of Neutron
Stars in Low Mass X-ray Binaries.
}
\author{  Yuri Levin
}
\affil{Theoretical Astrophysics, California Institute of
Technology, Pasadena, California 91125}
\authoremail{yurlev@tapir.caltech.edu}
\date{\today}
\begin{abstract}
Recently Andersson et. al., and Bildsten have independently
suggested that an r-mode instability might be responsible 
for stalling the neutron-star spin-up in strongly accreting,
Low Mass X-ray Binaries (LMXBs). We show that if this does 
occur, then there are two possibilities for the resulting
neutron-star evolution:

If the r-mode damping is a decreasing function of temperature,
then the star undergoes a cyclic evolution: (i) accretional 
spin-up triggers the instability near the observed maximum spin rate;
(ii) the r-modes become highly excited through gravitational-radiation-reaction,
and in a fraction of a year ($0.13$yrs in a particular model that we 
have considered) they viscously heat the star up to $T\sim 2.5\times 10^9$K;
(iii) r-mode gravitational-radiation-reaction then spins the star down in  
$t_{\rm spindown}\simeq 0.08 (f_{\rm final}/130\hbox{Hz})^{-6}$yrs to a limiting 
rotational frequency $f_{\rm final}$, whose exact value depends on the not
fully understood mechanisms of r-mode damping; (iv) the r-mode instability shuts off;
(v) the neutron star slowly cools and is spun up by accretion for $\sim 5\times 10^6$yrs,
until it once again reaches the instability point, closing the cycle. The shortness
of the epoch of r-mode activity makes it unlikely that r-modes are currently 
excited in the neutron star of any galactic LMXBs, and unlikely that 
advanced LIGO interferometers will see gravitational waves from extragalactic LMXBs.
Nevertheless, this cyclic evolution could be responsible for keeping the rotational
frequencies within the observed LMXB frequency range.

If, on the other hand, the r-mode damping is temperature independent, then a steady state
with constant angular velocity  and $T_{\rm core}\simeq 4\times 10^8$K is reached, in which
r-mode viscous heating is balanced by neutrino cooling and accretional spin-up torque
is balanced by gravitational-radiation-reaction spin-down torque. In this case (as 
Bildsten and Andersson et. al. have shown) the neutron stars in LMXBs could be potential
sources of periodic gravitational waves, detectable by enhanced LIGO interferometers.

\end{abstract}

\section{Introduction}
Most of the  rapidly accreting neutron stars in Low Mass X-ray Binaries (LMXBs) are observed
to rotate in a  strikingly narrow range of frequencies---from
$260$Hz to $330$Hz (see e.g.  Van der Klis 1997). A natural explanation for this 
could be some mechanism which  prevents  further neutron-star
 spin-up once the  rotational frequency is sufficiently high. Recently
several such mechanisms were proposed:

 White and Zhang (1997) 
suggested that magnetic braking could be responsible for halting the spin-up; 
 this idea will not be discussed here.
Bildsten (1998) pointed out that, because gravitational radiation reaction 
is a sharply increasing function of rotational frequency, it might halt the spin-ip.
In his original manuscript, Bildsten identified one mechanism for triggering the necessary 
gravitational waves: lateral density variations caused by temperature dependence of electron capture
reactions. While his manuscript was being refereed, Bildsten learned of the discovery that an
r-mode instability, driven by gravitational radiation, can be very strong  in spinning 
neutron stars  (Andersson 1998, Friedman and Morsink 1998,
Lindblom, Owen and Morsink 1998, Andersson, Kokkotas and Schutz 1998,
Owen et al 1998); and the r-mode experts learned of Bildsten's gravitational-wave idea
for saturating LMXB spinup. Both groups independently saw the connection:
Bildsten (1998) and Andersson, Kokkotas and Stergioulas (1998) proposed  that the
r-mode instability could provide enough gravitational-radiation reaction to halt
the LMXB spinup. In this letter we examine the consequences of this proposal.

 Our conclusions   depend crucially on whether 
the dissipation of the r-modes decreases with temperature (as is
the case, e.g., when shear viscosity dominates the r-mode damping),
or instead is temperature-independent (as is the case when, e.g., the 
mutual friction of proton and neutron superfluids dominates the damping). 
In the former case (Section $2$ of this paper) we 
find that the neutron star will undergo a spin-up---heating---spin-down---cooling
cycle; in the latter case (Section $3$) it will probably settle
down to a stable equilibrium state with an internal neutron-star
 temperature
of about $4\times 10^8$K.

\section{``Viscous'' r-mode damping}
Let us consider first the case when  dissipation is a decreasing
function of temperature. 
We show that, if some r-modes become unstable in a neutron star 
spun up by accretion, then they heat up the neutron star through 
shear viscousity. 
As the neutron star heats up, the r-modes become more unstable.
 A thermo-gravitational
runaway  takes place, in which the r-mode amplitude  grows, as
a result of this growth
the star's temperature rises, the dissipation becomes weaker
 and the instability becomes stronger. Within a fraction of
a year the r-modes' gravitational radiation reaction spins the star  down
to a rotation frequency which is close to the minimum
of the critical stability curve (probably around $100-150$Hz, but the
exact value depends on poorly understood dissipation mechanisms---see below), 
with a final
temperature of about $2\times 10^9$K. The instability then shuts off and the star
begins a several-million-year epoch of neutrino cooling
and accretional spinup, leading back to the original instability point.

Fig.\ 1 shows a typical evolutionary trajectory 
$A\rightarrow B\rightarrow C\rightarrow D\rightarrow B$ 
of the neutron star in the
$\log(T_8)-\tilde{\Omega}$ plane, where $T_8$ is the temperature of the
star's core
measured in units of $10^8$K, and
 $\tilde{\Omega}=\Omega/\sqrt{\pi G\bar{\rho}}$.
Here $\Omega$ is the angular velocity of the neutron star and $\bar{\rho}$
is it's mean density. 
The portion $A\rightarrow B$  of the curve represents the accretional
spin-up of the neutron star to the critical angular frequency 
$\Omega_{\rm cr}(T)$; $B\rightarrow C$ represents 
the heating stage in which the r-modes become unstable, grow and heat up the
neutron star;  $C\rightarrow D$ shows the spindown stage in which the r-mode amplitude
saturates because of poorly understood nonlinear effects,
and the angular velocity decreases due to the emission of 
gravitational radiation; and 
$D\rightarrow B$ represents cooling back to 
the equilibrium temperature with simultaneous spin-up by accretion.
 All four stages are discussed in more detail below.

 The initial (steady-state) temperature $T_0$ of the neutron-star
core in  steadily accreting LMXB's is somewhat uncertain; according 
to Brown and Bildsten (1998),  who analyzed  heat transport
during  steady thermonuclear burning of the accreted material and
nuclear reactions in the deep ocean, $T_0=1-4\times 10^8$K. In Fig. 1
we assume $T_0=10^8$K.

The curve $K-L-M$ is the so-called r-mode ``stability curve'' (Lindblom, Owen and Morsink 1998).
 If the neutron
star is represented by a point  above the  curve,
 then some  r-modes in the star are unstable and grow.
Otherwise, all r-modes decay.
The portion $K-L$ of the stability curve is determined by 
the shear viscosity,  or by  mutual friction if part of the star is
superfluid. Its exact location is uncertain precisely because
the dissipation of the r-modes  at the relevant temperatures
is poorly understood. If  shear viscosity
dominates the dissipation, then  the equation of the $K-L$
portion of the stability curve is given by
\begin{equation}
\tilde{\Omega}_{\rm cr}=0.1\left ({\eta\over \eta_0}\right )^{1/6}T_8^{-1/3},
\label{stabil}
\end{equation}
where $\eta$ is the shear viscosity of the neutron star material,
and $\eta_0$ is the shear viscosity due to electron-electron scattering in the
neutron star  [we have used Eqs\ (2.10), (2.14), (2.15) and Table I of
Owen et. al. (1998)  to work out Eq. (\ref{stabil})]. If 
only shear viscosity
due to to electron-electron scattering were operating,
with the shear viscosity given by $\eta_0=347\rho^{9/4}T^{-2}$, where
all quantities are in cgs units
 (see Cutler and Lindblom 1987 and references therein),  then the
critical rotational frequency at $T=10^8$K  would be $130$Hz, which is much less
than observed values (van der Klis 1998). However, the friction is probably 
larger than this (and therefore $\Omega_{\rm cr}$ is also larger)
because of interaction of the core fluid with the crust and maybe mutual
friction in a superfluid state. The emphasis of this
paper
is not to figure out whether the r-mode instability
is relevant for LMXBs, but to investigate the consequences if it is relevant.
For  purpose of illustration, 
we assume that $\eta=244\times\eta_0  $; this makes the critical rotational 
frequency  $330$Hz at $T=10^8$K, which is consistent with observations (van der Klis 1998). 
This choice of viscosity is a cheat since we don't yet know the $T$ and $\rho$-dependence
of $\eta$. However, unless the  damping
is due to  mutual friction, $\eta$ is likely
to decrease with increasing temperature,
 which is a  sufficient condition for thermo-gravitational runaway.
Our choice of  viscosity  possesses this feature; therefore we believe it has a good
chance of representing the real physics. 

The portion $L-M$ of the stability curve
is determined by bulk-viscosity dissipation;  it's exact location
is also a subject of yet unsettled controversy [the heart of the problem
is the calculation of Lagrangian perturbation in density (Lindblom, Owen and Morsink 1998,
Andersson, Kokkotas and Schutz 1998), which, to our best knowledge,
has not been reliably carried out by any of the groups]. Of the two current
estimates of 
 the  bulk-viscosity contribution to damping of the r-modes, we have 
chosen the one which gives the higher values of 
$\tilde{\Omega}_{\rm cr}$, thus maximizing its effect (see Lindblom, Owen and Morsink 1998).
The fact that, for the evolution curve shown
 in Figure\ 1
no part  is in the region where the bulk viscosity
dominates suggests that   the details of the bulk viscosity will
not  be of particular importance.

In this work  for concreteness we specialize to a polytropic model of a neutron star
with $p\propto \rho^2$, and consider the r-mode with $l=m=2$, which is expected
to have the strongest instability in such polytropes (Friedman and Morsink 1998,
Lindblom, Owen and Morsink 1998).
We assume that
the time evolution of the normalized  angular velocity $\tilde{\Omega}=\Omega/\sqrt{\pi G\bar{\rho}}$
 of the star and the dimensionless amplitude $\alpha$ of the 
r-mode are given by phenomenological Equations (3.14),  (3.15), (3.16) and (3.17) of Owen et al (1998):
\begin{equation}
{d\tilde{\Omega}\over dt}=-{2\alpha^2 Q\over 1+\alpha^2 Q}{\tilde{\Omega}\over\tau_v}
                           +\sqrt{4\over 3}{1\over\tilde{I}}{\dot{M}\over M}\times p,
\label{eq:omega}
\end{equation}
\begin{equation}
{d\alpha\over dt}=-\left({1\over\tau_{\rm grav}}+{1\over \tau_v}{1-\alpha^2 Q
\over 1+\alpha^2 Q}\right)\alpha
\label{eq:alpha}
\end{equation}
when $\alpha^2<k$ (the saturation value of $\alpha^2$, which we assume to be
$k=1$), and by
\begin{equation}
\alpha^2=k,
\label{eq:alpha1}
\end{equation}

\begin{equation}
{d\tilde{\Omega}\over dt}={2\tilde{\Omega}\over \tau_{\rm grav}}{kQ\over 1-kQ}
\label{eq:omega1}
\end{equation}
when $\alpha$ is saturated due to not yet understood non-linear effects.
Here $\alpha$ is the dimensionless amplitude of the r-mode defined by Eq.\ (1) of Lindblom,
Owen and Morsink (1998), and
$\tau_v$ and $\tau_{\rm grav}$ are the viscous and gravitational timescales for 
the r-mode dissipation and are given by Eqs (2.14) and (2.15) of Owen et al (1998):
\begin{eqnarray}
\tau_{\rm grav}&=&-3.26\tilde{\Omega}^{-6}\hbox{sec}       ,\nonumber\\
{1\over \tau_v}&=&{1\over\tilde{\tau_s}}\left({10^8\hbox{K}\over T}\right)^2+
                          {1\over\tilde{\tau}_ B}\left({T\over 10^8\hbox{K}}
                          \right)^6\tilde{\Omega}^2  .\label{eq:timescales}
\end{eqnarray}
In the above equation the viscous damping rate is a sum of contributions
from the shear and the bulk viscosities; the former is determined by
$\tilde{\tau_s}$ which we took to be $1.03\times 10^4$sec in order
to fit the observed data; the latter is determined by $\tilde{\tau}_B$ which
is taken to be $6.99\times 10^{14}$sec, in agreement with Owen et. al. (1998).

Note that $\tau_{\rm grav}$ is  negative  since gravitational 
radiation always amplifies the r-mode. The second term in Eq. (\ref{eq:omega})
represents the neutron-star spin-up by accretion; $M$ and $\dot{M}$ are 
the mass of the neutron star and its accretion rate respectively, and $p$
is a factor of order unity which depends on the accretion radius and
the angular velocity of the neutron star; it's exact value is not essential 
for the physics discussed here and we set $p=1$ from here onwards. 
 The numerical parameters $Q$ and $\tilde{I}$
are  given by $0.094   $ and $0.261  $ respectively for a polytrope star 
of adiabatic index $\gamma=2$ (Lindblom, Owen and Morsink 1998). For the 
evolution shown in Fig.\ 1 we took $M=1.4M_{\odot}$ and 
$\dot{M}=10^{-8}M_{\odot}/$yr, and we  assumed a random initial perturbation
of magnitude $\alpha=10^{-8}$ when the neutron
star reaches the stability curve $K-L$.

Now consider the star's thermal evolution. The r-mode deposits heat into the star
at the rate 
\begin{equation}
W_{\rm diss}={2E_{\rm r-mode}\over\tau_v}={\alpha^2\Omega^2 MR^2 \tilde{J}\over\tau_v},
\label{wdiss}
\end{equation}
 where $E_{\rm r-mode}$
 is the energy in the r-mode [cf. Eq. (3.11) of Owen et al (1998)].
Here $R$ is the radius of the neutron star taken to be $12.53$km, and $\tilde{J}=1.635\times 10^{-2}$
for the  polytropic model considered here.
At the relevant temperatures the neutron star is expected to cool predominantly by the modified URCA process
(this is not entirely true, since close to $10^8$K neutrino bremstruhlung cooling from the crust
and  radiative cooling by photons might become  significant. However, their cooling rates
are not significantly larger than that of the modified URCA process at $10^8$K, and they become negligible
at higher temperatures. In this work for simplicity we assume that  modified URCA
is the only cooling process; the inclusion of other processes would not change the general evolutionary 
picture).
The modified URCA cooling rate, reduced by heating from nuclear reactions in the
deep crust, is given by (Shapiro and
Teukolsky, 1983)
\begin{equation}
L_{\rm cool}=7\times 10^{31}(T_8^8-\bar{T}_8^8)\hbox{erg/sec}.
\label{lcool}
\end{equation}
Here
the subscript $8$ indicates that the temperature is measured in units of $10^8$K and
 $\bar{T}$ is the equilibrium 
temperature of the neutron star before the r-mode heating starts, taken to be $10^8$K for our calculation. 
The thermal evolution equation is then given by 
\begin{equation}
{dT\over dt}={W_{\rm diss}-L_{\rm cool}\over C_v},
\label{eq:temperature}
\end{equation}
where $C_v$ is the heat capacity of the 
 neutron star, taken to be $1.4\times 10^{38}$(erg/K)$\times T_8$ [from Shapiro and
Teukolsky (1983), Eq. (11.8.2). However, the heat capacity of neutron star with a superfluid
core is less]. 

Equations  (\ref{eq:omega}), (\ref{eq:alpha}), (\ref{eq:omega1}), (\ref{eq:alpha1})
 and (\ref{eq:temperature}) determine the time evolution of 
the angular velocity $\Omega$ and temperature $T$. Figure\ 1 shows the predicted
evolution, for the representative parameter values, introduced above. The evolution
consists of four stages:

 The first stage  $A\rightarrow B$ is  the  spin-up of   
the neutron star, during  which it's angular velocity $\Omega$ 
is increasing towards the 
critical one, and the r-mode instability is suppressed 
by viscosity; since we assume that the star begins at its 
equilibrium temperature $T_8=\bar{T}_8=1$, its
temperature changes little during the spin-up.
For an assumed accretion rate of $10^{-8} M_{\odot}/$yr this 
stage takes $\sim 5\times 10^6$ years. 

When the angular velocity reaches its critical
value, the r-mode starts to grow and 
 the second stage $B\rightarrow C$ begins.  The
neutron star gets heated up by the r-mode through viscosity, 
the r-mode becomes more unstable, and 
thermo-gravitational runaway  follows.
It takes $0.13$ years for the r-mode's amplitude to evolve from $\alpha=\alpha_{\rm W}$ to 
$\alpha=1$, where $\alpha_{\rm W}\simeq 1.2\times 10^{-5}$ is the value of the r-mode
amplitude at which the accretional torque is exactly compensated by the  
gravitational radiation reaction [see Wagoner (1984)].
For our intuition it is useful to define two characteristic $\alpha$-dependent timescales for stage
$B\rightarrow C$: 
the thermal timescale [cf Eq.\ (\ref{wdiss})]
\begin{equation}
t_{\rm th}={dt\over d\log T}={C_v T\over W_{\rm diss}}\sim 3.7\times 10^{-5}T_8^2{\tau_v\over \alpha^2}
\label{tth}
\end{equation}
and the timescale for the decrease of angular velocity [cf Eq.\ (\ref{eq:omega})]
\begin{equation}
t_{\Omega}={dt\over d\log\Omega}\simeq {1\over 2Q}{\tau_v\over \alpha^2}\simeq 5\times {\tau_v\over \alpha^2}.
\label{tomega}
\end{equation}
Clearly, the neutron star heats up much faster than it spins down due to gravitational radiation.
Therefore, during this stage the angular velocity of the star decreases by only a small
amount, $\Delta\tilde{\Omega}=0.0003$.
Physically, the reasons for such little change in $\tilde{\Omega}$ are that the
 r-mode amplitude grows so quickly, and that in this phase
the angular momentum loss  is not manifested
in a reduction of the angular velocity, but instead in  the growth of the 
r-mode itself (the r-mode which is driven by gravitational radiation reaction carries a negative angular
momentum).

 Eventually the r-mode
amplitude  saturates due to  nonlinear effects. 
This initiates the  third stage of the evolution,
in which all of the angular momentum loss is manifested by  reduction
of angular velocity (since the r-mode cannot grow any more), and the star spins down  
$C\rightarrow D$ to the critical angular velocity. At point $C_1$ the temperature
of the neutron star is such that the neutrino cooling exactly compensates the 
dissipative heating from the r-modes. After that the temperature does not change much until
the spin-down stage is terminated. The physical reason for this is that even though the thermal timescale
at $C_1\rightarrow D$ is comparable or smaller than the spindown timescale, the rate of dissipative heating
does not change much. If the heat capacity $C_v$ of neutron star were zero,
we would have $W_{\rm diss}=L_{\rm cool}$ at all points of $C_1\rightarrow D$. This would imply
$T\propto\Omega^{1/4}$, so even then the temperature would not change significantly over
this last part of the spin-down.
 
An analytical expression for the duration of this rapid spin-down
stage can be derived  from Eqs\ (\ref{eq:omega1})\ and\ (\ref{eq:timescales}):
\begin{equation}
t_{\rm spindown}\simeq 0.08(1/k)(\tilde{\Omega}_f/0.1)^{-6}\hbox{yr},
\label{down}
\end{equation}
where $\Omega_f$ is the angular velocity at the end of the spin-down.
In our simulations $t_{\rm spindown}$ is about $0.14$ years. 

After the neutron star reaches the stability curve,
 the r-mode is damped by viscosity stronger than 
it is driven by gravitational-radiation reaction; therefore its amplitude decreases
and the neutron star cools back to its original equilibrium temperature, while
being spun up by accretion. 
This part of the evolution is represented by $D\rightarrow B$ on 
 Fig.\ 1; its timescale is the same as that for the original accretional spin-up, i.e. $\sim 5\times 10^6$ years. 
After this the cycle is closed and can repeat itself as long as the accretion continues. 

We believe that the sharp kink at 
point $C$ is not
 a real physical effect, but a result of
 our poor understanding of the non-linear saturation of the r-mode;
however, this artificial feature of our
 simulations does not seem to affect the existence of the thermo-gravitational 
runaway and the subsequent rapid spin-down to a lower angular velocity.
Despite a large number of uncertainties in the details
of the  evolution, we believe that this scenario is
 robust so long as the r-mode instability does occur in LMXBs, and the
damping of the r-modes decreases with temperature. 

If the above described evolutionary scenario is generic, it is then clear that
none of the currently observed LMXBs can possess an actively operating
r-mode instability---otherwise we would observe a rapid spindown on a 
time-scale less than a year.  However, it is conceivable that many 
of the neutron stars in these LMXBs have  undergone the r-mode instability at some
stage of their evolution, and are currently below the stability curve, evolving along
leg $D\rightarrow B$ of Fig.\ 1.

From Equations (\ref{down}) and (\ref{eq:omega}) we can estimate the fraction $r$ of
neutron stars in extragalactic LMXBs that are  in the phase of active emission of
gravitational waves:
\begin{equation}
r={t_{\rm spindown}\over t_{\rm accretion}}\sim 1.6(1/k)\times 10^{-8}\left({\tilde{\Omega}_f\over 0.1}\right)^{-6}.
\end{equation}
The quantity $\tilde{\Omega}_f$ is bounded from below by the rotational frequencies
of young pulsars (this statement is true only if 
the r-mode damping is the same for young and old pulsars at the same temperatures).
 The rotational frequency of the recently discovered 
N157B (Marshall et al 1998)  is $62.5$Hz. Using the braking index theory
one can project the initial
rotational frequency of this pulsar to be no smaller then $100$Hz, which implies $\tilde{\Omega}_f>0.08$. 
Therefore, only $r<(6/k)\times 10^{-8}$
of neutron stars in extragalactic LMXBs  are in the phase of rapid gravitational wave emission,
which implies that to catch one star in this phase, gravitational-wave detectors must reach out 
through a volume large enough to encompass $\sim 0.1-0.01/r\sim 10^6$ galaxies like our own (this assumes
that there are $10-100$ strongly accreting neutron
stars in LMXBs in our galaxy). An analysis similar to that
of Owen et al (1998) shows that even ``advanced LIGO'' detectors are unlikely to be able to 
see these sources at such great distances.

\section{ temperature-independent r-mode damping}
There is  a possible  alternative  evolutionary scenario which is similar to the one
proposed by Bildsten (1998) and Andersson, Kokkotas and Stergioulas (1998)
(we thank Lee Lindblom for pointing this out). It may be that the r-mode damping 
is dominated not by normal dissipative processes, but by  mutual friction in
the neutron-proton superfluid. Detailed calculations of the effect of such
friction on the r-mode damping   are
in progress (Lindblom and Mendell);
 however for our analysis the essential feature of this dissipative process is
already known:  it is temperature
independent. Therefore, if this process dominates, one 
would not  expect a thermo-gravitational runaway;
instead the neutron star will reach a state of  three-fold equilibrium.
The neutron star will ``sit" on the stability curve [$(1/\tau_{\rm grav})+(1/\tau_v)=0$],
the amplitude of the r-mode will adjust so that the accretional torque
is compensated by the gravitational-radiation reaction torque ($\alpha=\alpha_W\simeq 1.2\times
10^{-5}$ for our model),
and the temperature of the neutron star will adjust so that the cooling  compensates
the frictional heating from the r-mode: $W_{\rm diss}=L_{\rm cool}$.
From Eqs\ (\ref{eq:omega}), (\ref{eq:timescales}), (\ref{wdiss}) and (\ref{lcool})
 one can work out the equilibrium temperature:
\begin{equation}
T_{\rm eq}=4.2\times 10^8\hbox{K}\left({f\over 330\hbox{Hz}}\right)^{1/8}\left({\dot{M}\over
  10^{-8}M_{\odot}/\hbox{y}}\right)^{1/8}\left({1.4M_{\odot} p\over M}\right)^{1/8},
\label{eq:equilibrium}
\end{equation}
where $f$ is the rotational frequency of the star.

It is interesting to examine how (and whether) the star reaches this equilibrium point.
For temperature-independent damping, Eqs\ (\ref{eq:omega}) and (\ref{eq:alpha})
form a closed system with two independent variables, $\tilde{\Omega}$ and
$\alpha$. To investigate the behavior
of the star after it reaches the stability curve at $\tilde{\Omega}=\tilde{\Omega}_{\rm cr}$,
we set $\tilde{\Omega}=\tilde{\Omega}_{\rm cr}+\tilde{\Omega_1}$ and expand 
Eqs\ (\ref{eq:omega}) and (\ref{eq:alpha})  to  first order in $\tilde{\Omega_1}$.
After trivial algebraic manipulations, we can then reduce the system 
of two first-order differential equations to a single second-order differential equation:
\begin{equation}
{d^2 x\over d t^2}+\gamma(x){dx\over dt}=-{\partial V(x)\over \partial x}.
\label{particle}
\end{equation}
Here $x=\ln \alpha$, and $\gamma (x)$ and $V(x)$ are given by
\begin{equation}
\gamma (x)={2Q\exp (2x)\over \tau_v}
\label{gamma}
\end{equation}
and
\begin{equation}
V(x)={6\over \tau_{\rm grav}}\left({Q\exp (2x)\over \tau_v}-{x\over\tau_{\rm acc}}\right).
\label{V}
\end{equation}
In the above Equations $\tau_{\rm grav}$ is given by Eq. (\ref{eq:timescales}),
and $\tau_{\rm acc}=(1/p)\sqrt{(3/4)}\tilde{\Omega}_{\rm cr}\tilde{I}M/\dot{M}$ is the timescale
for the neutron star to be spun up by accretion to the angular frequency $\Omega_{\rm cr}$.

Clearly Eq.\ (\ref{particle}) can be thought of as an equation of motion for a particle
of unit mass in the potential well given by $V(x)$ and with the damping $\gamma(x)$.
The bottom of the potential well corresponds to the equilibrium state described above,
 and the damping
insures that the ``particle'' gets there (i.e. that the neutron star settles into the equilibrium state).
However, the damping is small. To see this, consider damped oscillatory motion
close to the bottom of the  well. The complex angular frequency of this
motion is given by
\begin{equation}
\omega=\sqrt{12/( \tau_{\rm acc}\tau_{\rm grav})}-i/(2\tau_{\rm acc}).
\label{omega}
\end{equation}
The period of these small oscillations is 
\begin{equation}
P\sim 230 \left({M\over 1.4 M_{\odot}}\right)^{1/2}
\left({10^{-8}M_{\odot}\hbox{y}^{-1}\over\dot{M}}\right)^{1/2}\left({f\over 330\hbox{Hz}}\right)^{-5/3}\hbox{y},
\label{P}
\end{equation}
but the timescale on which they are damped (i.e. the
timescale on which the equilibrium is reached) is $\tau_{\rm eq}\sim 2\tau_{\rm acc}\sim 
10^7$y.   

Since the damping is so small, fluctuating disturbances may keep this
nonlinear oscillator off it's equilibrium position. For example, in our
evolutionary scenario we
have  assumed that there is a mechanism which gives $\alpha$ some
non-zero initial value. Presumably, the same mechanism could keep the oscillator
in an excited state. Then the amplitude
of the r-mode, and hence the temperature of the star's core, would vary
on the timescales of hundreds of years.  Detailed investigation of these issues
is a subject for further work. However it is clear that the time-averaged
temperature should be close to the equilibrium value given by Eq.\ (\ref{eq:equilibrium}).
 
If the r-mode damping does not depend on temperature, we can expect r-modes
to be excited in many of the rapidly rotating neutron stars in
LMXBs.
These presumably superfluid
steady gravitational-wave emitters could be detected by enhanced LIGO
gravitational wave detectors, as   discussed in Bildsten (1998) and
 Andersson, Kokkotas and Stergioulas (1998). Recently, Brady and Creighton (1998) have considered
the computational cost of such detection. Their conclusion was that with the enhanced LIGO
sensitivity and available computational capabilities one could detect gravitational-wave emitters in
LMXBs that are as bright in X-ray flux as SCO-X1. 

If the rotational frequency of the emitting neutron star 
is  localized to within a few 
$10$s of Hz using astronomical observations (by, e.g., QPO's), one could narrow-band 
the interferometer response around the frequency of r-mode oscillations (see e.g. Meers 1988).
 This could
allow LIGO to detect gravitational-wave emitters in LMXBs which are $10-100$ times dimmer in
X-ray flux than SCO-X1.

Positive detection of gravitational waves at the r-mode oscillation frequency would
make a strong case for 
 the superfluid nature of the r-mode damping.

\section{Conclusions}
In this paper we have investigated the recent proposal that the accretional 
spin-up of the neutron star in an LMXB is stopped by r-mode gravitational radiation 
reaction. There are two possible evolutionary scenarios. 
In the first scenario, the neutron star  goes through cycles such as that shown
on Fig.\ 1. The necessary condition for this
scenario to be relevant is that r-mode damping
should decrease with increasing temperature.
 In this case, it is very unlikely that any of the currently observed 
neutron stars in LMXBs in our galaxy are
in the r-mode excited phase of the cycle.  The detection
of gravitational radiation from extragalactic LMXBs in 
the r-mode excited phase  is also not likely, even with advanced LIGO
interferometers.

 In the second scenario, r-mode damping is temperature independent, and a steady-state
equilibrium is probably reached, where both angular velocity and temperature stay constant,
or are oscillating with periods of several hundreds of years.
Equation (\ref{eq:equilibrium}) makes a robust prediction for the temperature 
of these objects to be $\simeq 4\times 10^8$K; this temperature is on the high 
end of what is  typically expected;
and it might be possible to test this prediction
by observations.
In this  case the neutron stars are emitters of periodic gravitational waves, which could
be detected by interferometers like enhanced LIGO.

I wish to thank Kip Thorne, Lars Bidsten and Lee Lindblom  for making
suggestions and comments which are crucially important for this work.
 I thank Teviet Creighton, Curt Cutler,  Benjamin Owen, Sterl Phinney and
Bernard Schutz for discussions, and AEI-Potsdam, where part of this 
paper was completed, for hospitality. This work was supported by  NSF grants
AST-9731698 and PHY-9424337.

\newpage
\figcaption{Cyclic evolution of a strongly accreting neutron star in a LMXB. R-mode damping is
            assumed to decrease with the  temperature $T$ of the neutron-star core. Line $K-L-M$
            represents the ``stability curve''; when the neutron star gets above this line,
            r-modes grow due to gravitational radiation reaction. Leg $A\rightarrow B$ of
            the evolutionary track 
            represents the accretional
            spin-up of neutron star to the critical angular frequency;  $B\rightarrow C$ represents
          the heating stage in which the r-modes become unstable, grow and heat up the
           neutron star; $C\rightarrow D$ shows the spindown stage in which
            the angular velocity decreases due to the emission of
           gravitational radiation;  and
          $D\rightarrow B$ represents the neutron-star cooling back to
          the equilibrium temperature with simultaneous spin-up by accretion, thus closing the cycle.}
              

\newpage
\begin{references}
 \reference{andersson1} Andersson,\ N. 1998, gr-qc/9706075 [to appear in \apj]
 \reference{andersson2} Andersson,\ N., Kokkotas,\ K\. D.
   \& Stergioulas,\ N. 1988, astro-ph/9806089
 \reference{andersson3} Andersson,\ N., Kokkotas,\ K\. D. \& Schutz,\ B.\ F., astro-ph/9805225
 \reference{bildsten} Bildsten, L., \apj, 501, L89
 \reference{creighton} Brady,\ P.\ R. \& Creighton,\ T.\ D. 1998, in preparation
 \reference{brown} Brown, E.\ F.\ \& Bildsten, L. 1998, \apj, 496, 915
 \reference{cutler} Cutler,\ C. \& Lindblom,\ L. 1987, \apj, 314, 234
 \reference{friedman} Friedman,\ J.\ L. \& Morsink,\ S. 1998, gr-qc/9706073 [to appear in \apj]
 \reference{lindblom1} Lindblom,\ L. 1995, \apj, 438, 265
 \reference{lindblom2} Lindblom,\ L. \& Mendell,\ G. 1995, \apj, 444, 804
 \reference{lindblom3} Lindblom,\ L. \& Mendell,\ G. 1998, in progress
 \reference{lindblom4} Lindblom,\ L., Owen,\ B.\ J. \& Morsink,\ S. 1998, Phys. Rev. Lett., 80, 4843
 \reference{marshall} Marshall,\ F.\ E. et al 1998, \apj, 499, L179
 \reference{meers}  Meers,\ B.\ J. 1988, Phys. Rev. D, 38, 2317
 \reference{mendell} Mendell,\ G. 1991, \apj, 380, 530
 \reference{owen} Owen,\ B.\ J.  et. al. 1998, gr-qc/9804044, to be published in 
                 Phys. Rev. D 
 \reference{shapiro} Shapiro,\ S.\ L. and Teukolsky,\ S. 1983, Black Holes, White Dwarfs and Neutron
                    Stars (John Wiley and sons, New York)
 \reference{vanderklis} van der Klis, M 1998, in Proc. NATO/ASI Conf. Ser. C,
               The many faces of neutron stars, ed. R.\ Buccheri, J. van Paradijs and
                M.\ A.\ Alpar (Dordrecht: Kluwer), in press (astro-ph/9710016)
 \reference{wagoner} Wagoner,\ R.\ V. 1984, \apj, 278, 345
 \reference{white}  White, N.\ E.\ \& Zhang, W. 1997,  \apj, 490, L87
\end{references}
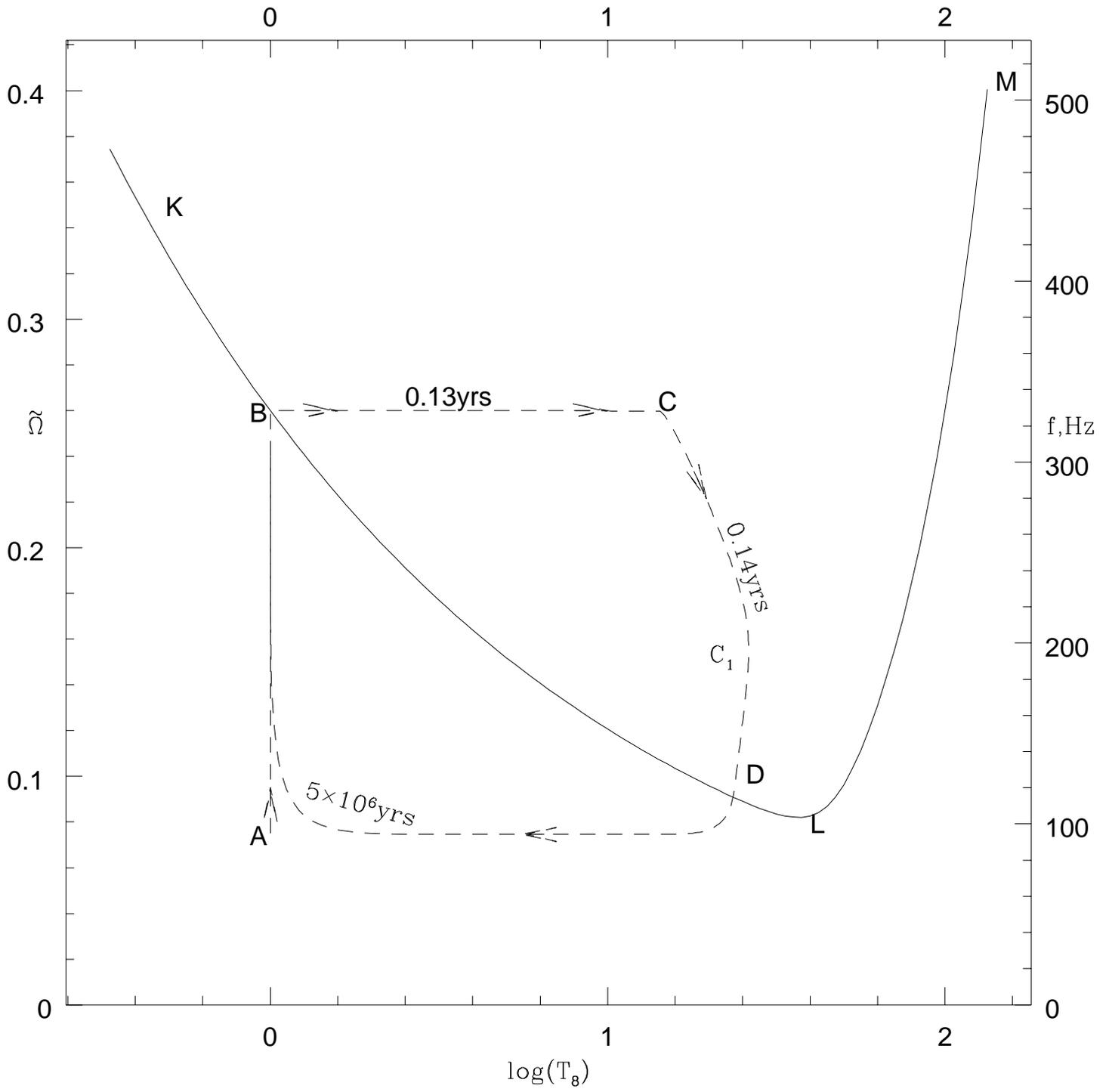
\end{document}